# On superconducting niobium accelerating cavities fired under $N_2$-gas exposure

Ralf Eichhorn[1], Daniel Gonnella[1], Georg Hoffstaetter[1], Matthias Liepe[1] and Wolfgang Weingarten[2]

[1] Cornell Laboratory for Accelerator-Based Sciences and Education, Ithaca, NY

[2] Visiting Fellow at Cornell University, Senior Physicist from CERN, Geneva (Switzerland), now retired

Abstract:
The dependence of the Q-value on the RF field (Q-slope) for superconducting RF cavities is actively studied in various accelerator laboratories. Although remedies against this dependence have been found, the physical cause still remains obscure. A rather straightforward two-fluid model description of the Q-slope in the low and high field domains is extended to the case of the recently experimentally identified increase of the Q-value with the RF field obtained by so-called "N-doping".

*Introduction*

The main thrust of research on superconducting (sc) accelerating cavities went into the reduction of the residual losses and also into achieving a high accelerating field gradient. The effort was a success and consisted mainly in a strict choice and processing of the niobium metal (high thermal conductivity metal, chemical surface processing, annealing, and high pressure water cleaning). The surface resistance could be reduced down to a few $n\Omega$. However, two different observations emerged: Either the Q-value increases with the field at very low gradients (low field Q-increase) or the Q-value goes down, gradually (Fig. 1(b), medium field Q-slope) or even sharply, particularly at the highest gradients (high field Q-slope or Q-drop). Solutions to lessen the Q-slopes were found by experiment [1] and consisted mainly of electro-polishing and bake-out at 120 °C for an extended period (~ days).

Recently, so-called "N-doped" niobium sc cavities obtained increased interest, because they hold the promise of large Q-values at technically still useful accelerating gradients. "N-doped" means that the cavity is heat treated in a nitrogen atmosphere (usually 20 - 50 mTorr) at high temperatures (usually 800°C for several tens of minutes). They exhibit an increase of the Q-value with the magnetic surface field B (negative Q-slope) up to a maximum field about 60 - 80 mT, equivalent to 15 - 20 MV/m accelerating gradient. This observation was repeatedly observed in different laboratories [2, 3, 4, 5], such as Cornell [6] (Fig. 1).

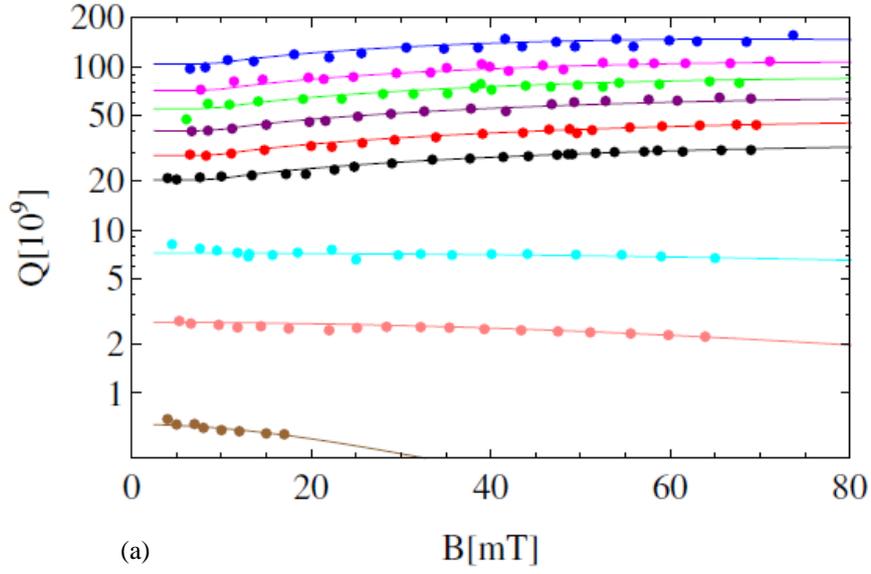

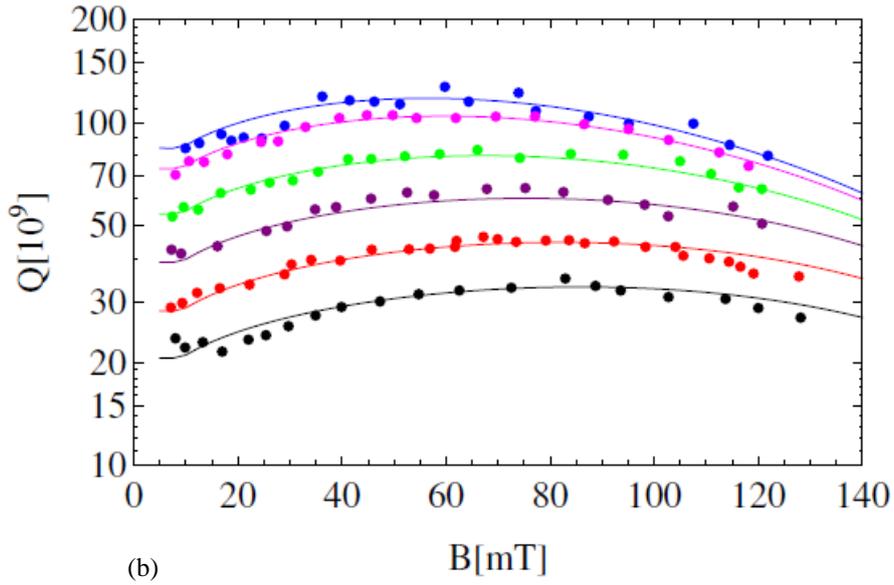

Fig. 1: (a) Q-value vs magnetic surface field B in 1.3 GHz single cell cavity, made of bulk niobium and fired in a $N_2$ atmosphere [p = 50 mTorr (66 hPa), 800°C, 20 minutes], electro-polished, prepared at FNAL and tested at Cornell at different temperatures; temperatures are, from top to bottom, 1.6, 1.7, 1.8, 1.9, 2.0, 2.1, 2.5, 3.0 and 4.2 K

(b) Q vs B curve of a similar cavity, submitted to a slightly different treatment (800°C in vacuum plus 20 minutes in 60 mTorr of $N_2$ followed by 30 minutes in vacuum) and electro-polished (13μm); temperatures are, from top to bottom, 1.6, 1.7, 1.8, 1.9, 2.0 and 2.1 K.

For an accelerating gradient of 10 MV/m the surface magnetic field B amounts to 42 mT. The continuous lines result from the fit to the data with $Q = 270.7 \, \Omega/R_s$, with $R_S$ being functionally described in the text. The gradient in (a) was limited by a quench, no limitation was observed in (b).

This observation has practical implications for the operation of a particle accelerator, having as most welcome consequence a reduction of the cryogenic load. It is also promising with respect to a better understanding of the field dependence of the Q-value in general. Some practical recent applications of the N-doping recipe are found in ref. [7].

Nevertheless, accelerating cavities treated by N-doping are not free from the previously observed decrease of the Q-value with the RF magnetic field. This is, particularly at the highest accelerating gradients, an undesirable effect for accelerator application.

These observations were described in several models [8, 9, 10], but a unanimously accepted explanation is missing. A better insight would give hope to improve the already applied practical measures to increase the Q-value and so improve the operation of sc accelerating cavities in large particle accelerators or other applications. The present paper presents a model that should contribute to the explanation of the observations and presents a common description of the observed Q-increase and Q-decrease.

Our model states that the surface layer is non-uniform in terms of having defects being only weak superconductors, as outlined in ref. [9], and extends it to deeper lying defects inside the bulk. A weak sc defect is supposed of mesoscopic size, embedded in the "host" sc metal of high purity and in close contact with this. The defect has sc properties (Cooper pairs, characteristic coherence length $\xi_N$) induced from, and weaker than the "host" sc metal Nb (lower critical field and temperature of the defect compared to the host metal). The defect also induces nc charge carriers into the otherwise sc host metal. All these features are typical of the sc proximity-effect described elsewhere [11]. Furthermore, when the defect is in the nc state, the current across the nc - sc interface may be impeded by the lack of states within the sc energy gap at the Fermi level.

We will treat these defects as an additional resistance but assume that surface currents are able to divert them. Interestingly, this allows describing the increase in Q (and decrease in losses) with increasing RF current. The proposed model is general and should be valid for various sc cavities of a large variety, those used for ultra-relativistic beams (ß = 1) structures as well as and those used for low-ß structures.

*Spirit of the present data analysis*

Our analysis is based on the two-fluid model of RF superconductivity in the London formulation with emphasis on the electrical conductivity of the normal-conducting (nc) component of the superconductor. We often make no difference between the electrical conductivity attributed to the nc component and the number of nc electrons (quasiparticles), because they are proportional to each other.

We presume that alternate attempts to explain the field dependence of the Q-value are valid, but, as we believe, so far in weaker agreement with the published data, and are therefore not dealt with. These are, for instance firstly, the temperature build-up created by the heat flow across the cavity wall (thermal feedback model) mediated by the exponential temperature dependence of the BCS surface resistance [8, 12]. Secondly, the superconducting (sc) energy gap may depend on the magnetic field B or the surface current [8]. The reasons for neglecting these attempts are, for the first case, the much stronger dependence of the Q-value with the field as observed in micron-thin film niobium cavities, the walls of which are backed by high thermal conductivity copper as for the LEP and LHC colliders at CERN [13]. The field dependence in that case should be much smaller than in bulk niobium cavities, the contrary of which is observed. For the second case it is difficult how to combine the proposition of a field dependent energy gap with the often observed factorization of the Q vs B curves into a field dependent and a temperature dependent part [14].

We use the method as described previously [9] and extend it. It consists of a rather straightforward thermodynamic energy balance consideration to estimate the critical field and the nc volume increase of weak sc defects associated with the RF magnetic field B, both at the surface and, what follows, in the bulk as well. Our approach is mostly phenomenological and does not go further into microscopic elaboration. It should allow therefore a better insight into the underlying physics, possibly somewhat at the cost of attention to scientific detail. In ref. [9], the critical temperature was determined, as a showcase, in relation to a composite made up of a strong superconductor (Nb) and a normal conductor in its proximity (NbO above 1.4 K). A percolation threshold for the composite was identified for a weak sc defect at the surface. We extent this approach for a weak sc defect located in the bulk and subject to the sc proximity effect, however no longer for the specific case of a Nb/NbO composite. Finally we check these ideas on the data of "N-doped" cavities, as explained and depicted in Fig. 1.

*The critical field and the nc volume associated with the RF magnetic field B*

*Weak sc defect at the surface*

Energy balance considerations are resumed as outlined in ref. [9]. They lead to the following considerations: If the magnetic RF field B is raised, the superconductor will gain diamagnetic volume energy $E_m$ by allowing of magnetic flux, not in the form of vortices, however, to penetrate into the magnetic volume $V_m$ at the surface in the vicinity of a weak sc defect: $E_m = B^2/(2\mu_0) \cdot V_m$. The penetration would continue if not a counteracting effect would stop it. The concomitant creation of a nc volume $V_c$ costs condensation volume energy $E_c = B_c^2/(2\mu_0) \cdot V_c$. Therefore the energy balance requests $E_m = E_c$, or, equivalently,

$$B_c^2 \cdot V_c = B^2 \cdot V_m \qquad . \qquad (1)$$

Eq. (1) allows two conclusions. The first one points towards a small first entry of magnetic field at a point-like surface weak sc defect with dimensions considered small as compared to the characteristic lengths inside a sc metal in its vicinity, the coherence length $\xi$ and the penetration depth $\lambda$. This special spherical geometry permits magnetic flux to penetrate at and above a magnetic field $B_c^*$, if, according to eq. (1), $B_c^{*2} \cdot \lambda^3 = B_c^2 \cdot \xi^3$, hence

$$B_c^* \approx \frac{B_c}{\kappa^{3/2}} \quad , \tag{2}$$

with the Ginzburg-Landau parameter $\kappa = \lambda/\xi$. The local critical field $B_c^*$ can be much smaller than the thermodynamic critical field $B_c$, because the local coherence length $\xi$ close to the surface may become pretty small and the local penetration depth at the surface rather large. For a mean free path $l = 10$ nm, for example, one finds for niobium $\xi = 8$ nm, $\lambda = 83$ nm and $B_c^* = 6$ mT (with $B_c = 190$ mT).

The second conclusion from eq. (1) concerns the spatial development of the nc region where magnetic flux enters for a magnetic field B above $B_c^*$. As the condition as of eq. (1) still holds, the increase of the condensation volume $\Delta V_c$ is

$$\Delta V_c = \frac{2B \cdot \Delta B \cdot V_m}{B_c^2} + \frac{B^2}{B_c^2} \Delta V_m \; .$$

As outlined in ref. [9], for the nc volume still small compared to both $V_c$ and $V_m$, the volume $V_c$ and consequently the electrical conductivity $\sigma$ increases with B as

$$\sigma \sim -\frac{1}{\kappa^2}\left\{1 + \frac{\ln\left[1-\left(\frac{\kappa B}{B_c}\right)^2\right]}{\left(\frac{\kappa B}{B_c}\right)^2}\right\} = \frac{1}{\kappa^2}\left\{\frac{\left(\frac{\kappa B}{B_c}\right)^2}{2} + \frac{\left(\frac{\kappa B}{B_c}\right)^4}{3} + \cdots\right\}. \tag{3}$$

Eq. (3) describes the medium field Q-slope and, by force of the singularity at $B = B_c/\kappa$, the high field Q-drop as well.

These considerations apply for a weak sc defect at the surface. That the surface quality is important with regard to the dependence of the Q-value with the RF field was already observed before, when a 1.5 GHz niobium thin film cavity exhibited a largely reduced dependence after a high pressure water rinsing [15], which is considered to involve the surface but not the bulk.

*Weak sc defect in the bulk*

The new data of an N-doped cavity (Fig. 1) allow describing how a deep-lying weak sc defect located in the bulk will react to the RF magnetic field. Considering this case is obvious, because the N-doped cavities show best performance after chemical removal of several μm of niobium. In addition, the SIMS elemental depth profile shows large excess nitrogen at the surface and down into the bulk (Fig. 2), however less there, corroborated elsewhere, too [16].

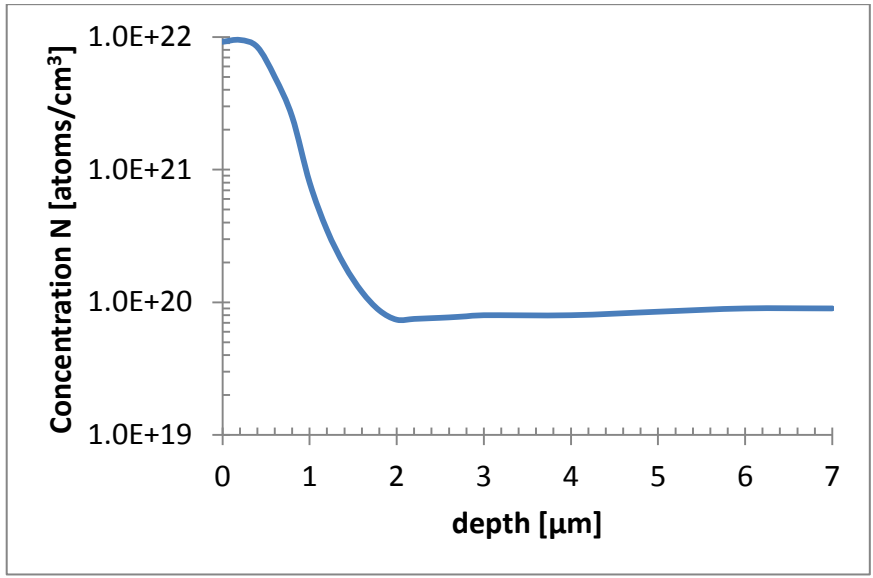

Fig. 2: SIMS depth profile of the nitrogen content for a $N_2$-doped surface.

For a weak sc defect in the bulk, exposed to the RF magnetic field B, there is no gain in diamagnetic volume energy with increasing B. This is visualized in Fig. 3. The RF current distribution in the vicinity of nc defects (i.e. the weak superconductor above its critical field) in the current-carrying layer is shown for both cases, if located at the surface or if located in the bulk. The underlying proposition holds that the RF current avoids to a large extent the nc defect and circumvents it, as in the static case. We convinced ourselves that this approach is justified, because the electromagnetic field differential equations as applied inside the metal come close to the static or low frequency case for the range of frequency and electrical conductivity applicable here. Then, the current in the nc state (upper right) may be thought of a superposition of the current for the two defects still being in the sc state (upper left) and of two dipole toroid-like current configurations (upper middle). The second and third lines depict the magnetic field. It is evident that there is a decrease of the diamagnetic volume for the defect at the surface (middle right), but none for the defect in the bulk (lower right).

The weak sc defect is also subject to the sc proximity effect, being in the sc state thanks the proximity of a strong superconductor, otherwise being nc. The sc proximity effect is characterized by a relatively small critical magnetic field $B_c^*$ (in the 10 mT region), depending on the size of the weak sc defect. At very small RF magnetic fields, the weak sc defect remains still sc until the local RF magnetic field B inside the metal surpasses its critical magnetic field $B_c^*$ there. If B is even more increased, those weak sc defects located within the distance $z_c$, for which B inside the metal $B_z = B \cdot \exp(-z/\lambda)$ exceeds $B_c^*$, are nc (Fig. 4). The distance $z_c$ of weak sc defects having already transited into the nc state increases therefore with B according to $z_c(B) = \lambda \cdot \ln(B/B_c^*)$. So does the volume fraction f(B) of the weak sc defects when in the nc state. The provision is made that the repartition of the weak sc defects is uniform with depth z. This is, according to Fig. 2, to a large extent correct after removal of the surface layer by electro-polishing.

In what follows we derive the sc surface resistance $R_s$ of a slab as depicted in Fig. 4. For convenience we apply the lumped circuit model description of current flow.

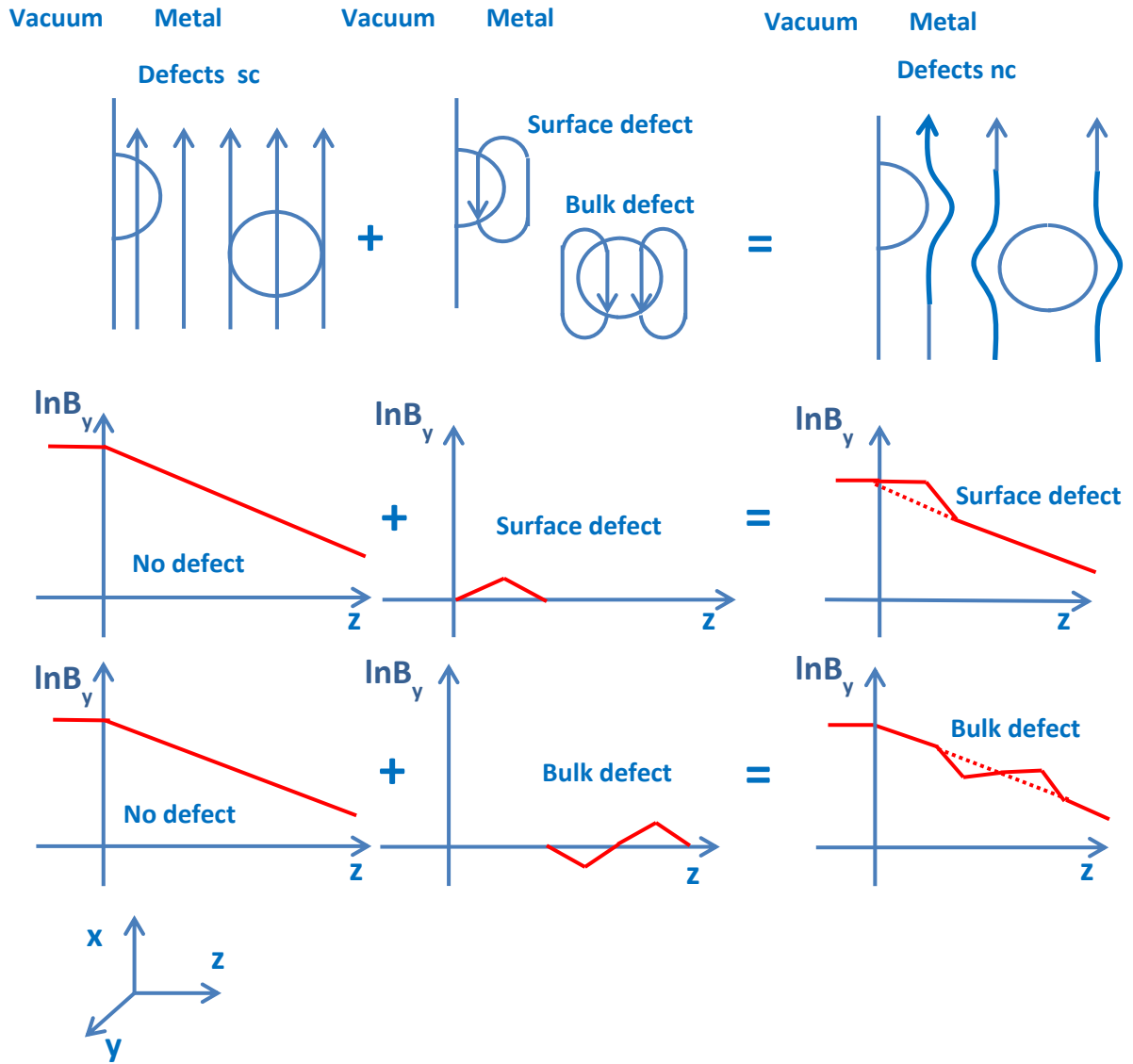

Fig. 3: Visualization of the sc current distribution around nc defects at the surface and in the bulk.

Inspecting Fig. 4, the current flows along the x-axis, and is driven by the electric field E. The power dissipation $\Delta P$ inside the slab of width $\Delta x$, cross section $\lambda \cdot l$, and resistance R is

$$\Delta P = V^2/(2R). \qquad (4)$$

V is the voltage across this slab, which is created by the Meissner current at the surface, or, equivalently, the RF surface magnetic field B,

$$V = -i\omega\lambda B \cdot \Delta x. \qquad (5)$$

As prescribed by the elementary two-fluid model of superconductivity, the resistance R is composed in parallel circuit arrangement of the resistances $R_1$ and $R_2$,

$$1/R = 1/R_1 + 1/R_2, \qquad (6)$$

$R_1$ being associated with the super-current's nc component and $R_2$ being associated with the current across the weak sc defects, when they are nc. The corresponding conductivities are $\sigma_1$ and $\sigma_2$.

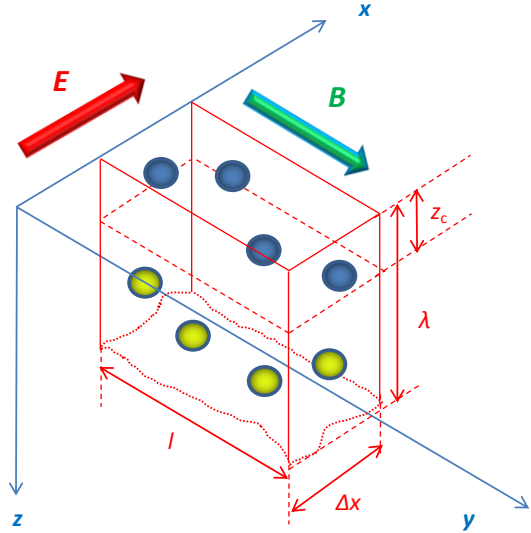

Fig. 4: View of current-carrying element: The surface current flows in x-direction, the magnetic and electric fields B and E decay exponentially deeper inside the metal with the characteristic decay length (the sc penetration depth λ). Minute weak sc defects are symbolized by colored spheres. A characteristic depth $z_c(B)$ depends on the surface magnetic field B and discriminates the upper part against the lower one. The weak sc defects located in the upper part (blue) are nc, whereas those located in the lower part (green) are still sc.

Now we assume that weak sc defects of size small compared to the sc penetration depth λ are located inside a surface layer (symbolized in Fig. 4 by the blue and green little spheres).

As outlined before, the volume fraction f of weak sc defects that already have become nc is zero for $B \leq B_c^*$ and increases for $B > B_c^*$ up till $B_c$, at the utmost, as

$$f(B) = \begin{cases} ln(B/B_c^*)/ln(B_c/B_c^*), & B \geq B_c^* \\ 0 & , else \end{cases} \quad , \tag{7}$$

with $f(B_c) = 1$. With the symbols as shown in Fig. 4 the respective resistances are

$$1/R_1 = (\lambda l/\Delta x) \cdot \sigma_1(T) \cdot [1 - f(B)] \tag{8}$$

and

$$1/R_2 = (\lambda l/\Delta x) \cdot \sigma_2 \cdot f(B). \tag{9}$$

Substituting eqs. (5) to (9) into eq. (4), we get with $B = -\mu_0 \cdot H$,

$$\Delta P = (1/2)\omega^2 \lambda^2 \mu_0^2 H^2 \cdot \lambda l \cdot \{\sigma_1(T)[1 - f(B)] + \sigma_2 f(B)\} \cdot \Delta x. \tag{10}$$

Therefore the electrical conductivity σ, averaged over the volume, is composed of a sum[a]. It consists firstly of the conductivity corresponding to the temperature dependent nc current component in the sc state $\sigma_1(T)$ (named $\sigma_n$ in ref. [9]), reduced by a volume fraction f(B) of weak sc defects that have already become nc. It consists secondly of the corresponding conductivity $\sigma_2$ of weak sc defects already in the nc state filling the same volume fraction f(B), being independent of T, and both $\sigma_1(T)$ and $\sigma_2$ of the same order of magnitude in the interesting temperature range. Hence we tacitly assume that the total volume fraction of the sc defects is constant, part of it being "dormant" and sc, the other part nc, with the splitting depending on B:

$$\sigma(T,B) \sim \sigma_1(T) \cdot [1 - f(B)] + \sigma_2 \cdot f(B) \quad . \tag{11}$$

The dissipated power per unit-square is $p=\Delta P/(l \cdot \Delta x)$, from which the surface resistance $R_s = 2 \cdot p/H^2$ is derived as

$$R_s = \omega^2 \lambda^3 \mu_0^2 \{\sigma_1(T)[1 - f(B)] + \sigma_2 \cdot f(B)\} = \overbrace{\omega^2 \lambda^3 \mu_0^2 \sigma_1(T)}^{R_{s,BCS}} \cdot \left[1 - f(B) + \frac{\sigma_2}{\sigma_1(T)} \cdot f(B)\right]. \tag{12}$$

However, we know that for $B \leq B_c^*$, when the expression in the bracketed parenthesis is 1, the surface resistance $R_s$ is composed of the BCS surface resistance $R_{s,BCS}$ and the residual surface resistance $R_{res}$, which we add and write in the usual way as

$$R_s = \left( \overbrace{A \cdot \frac{e^{-\Delta/T}}{T}}^{R_{s,BCS}} + R_{res} \right) \cdot \left[ 1 - f(B) + \frac{\sigma_2}{\sigma_1(T)} \cdot f(B) \right] \qquad . \qquad (13)$$

---

[a] In this derivation we neglect polarization effects due to the different conductivities $\sigma_1$ and $\sigma_2$. We convinced ourselves that no difference exists between our approach (eq. 11) and the more refined one as described by R. Landauer, The Electrical Resistance of Binary Metallic Mixtures, Journ. Appl. Phys. **23** (1952) 779.

A is a material and frequency dependent constant, Δ is the energy gap of the superconductor and $R_{res}$ describes temperature independent residual losses.

*Critical temperature of a nc/sc composite and percolation effects*

We digress now from examining a composite of a mixture of niobium and niobium-(mon)oxide as in ref. [9] except for applying the same physics, such as proximity effect and percolation. The proximity between the nc and sc metal induces nc charge carriers into the latter, thus increasing the normal state conductivity of the sc metal. By percolation the nc metal is fragmented with increasing temperature into smaller pieces separated by long range sc paths. Hence this fragmentation enlarges the proximity between nc and sc metal and increases the normal state conductivity of the sc metal once again.

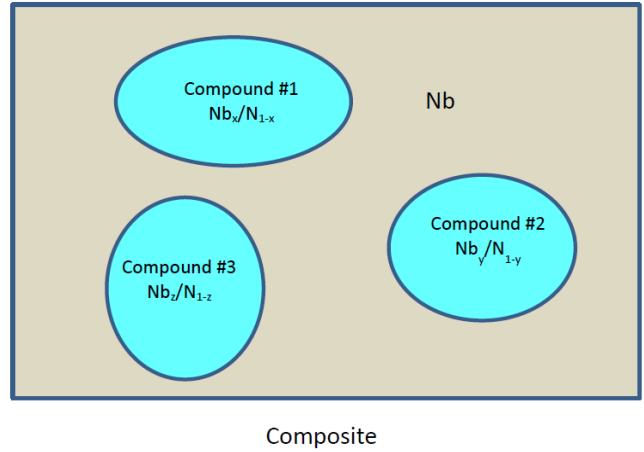

Fig. 5: Visualisation of "compound" and "composite".

We call now the weak sc defect composed of nitrogen dissolved in niobium ($Nb_x/N_{1-x}$) of still unspecified and variable atomic composition x of Nb a "compound". We call these compounds, if embedded in a niobium matrix, a "composite" (Fig. 5). If the composite is located at the surface, the nc volume will increase and is subject to the percolation action. The related surface resistance is therefore proportional to $\sigma_s$, which we call the conductivity of the surface,

$$\sigma_s \sim \Theta(T - T^*) \cdot (T - T^*)^\beta \qquad . \qquad (14)$$

$T^*$ is the "percolation temperature" above which the composites have obtained long-range connectivity and hence sharply increased electrical conductivity. The exponent β is a phenomenological exponent describing the increase of the nc volume with temperature above T*, and Θ is the Heaviside step function. If the weak sc defect is located in the bulk, the conductivity will increase stepwise above T'*, possibly different from T*, due to local percolation, to a constant value, as suggested by the constant N-depth profile, Fig. 2. We call the related electrical conductivity $\sigma_{bulk}$, being proportional to

$$\sigma_{bulk}(T,B) \sim \sigma_1(T) \cdot [1 - f(B)] + \sigma_2(T) \cdot f(B) \quad , \quad (15)$$

with

$$\sigma_2(T) = \sigma_{20} + \Theta(T - T'^*) \cdot \Delta\sigma_2 \quad .$$

The conductivities $\sigma_1$ and $\sigma_2$ are defined as in eqs. 8 and 9 except for the stepwise increase $\Delta\sigma_2$ of the conductivity $\sigma_2$ at T'*, otherwise being independent of the temperature. Above T'*, consequent to percolation, assumed to be complete, $\sigma_2$ should be equal to $\sigma_1$, hence $\Delta\sigma_2 = \sigma_1(T) - \sigma_{20}$ and $\sigma_{bulk}(T, B) = \sigma_1(T)$.

Table 1: Electrical conductivity σ for a weak sc defect located in the bulk or at the surface, resp.

| Location of weak sc defect: | Bulk | Surface |
|---|---|---|
| σ vs magnetic field B | $\sigma_{bulk}(T,B) \sim \sigma_1(T) \cdot [1 - f(B)] + \sigma_2 \cdot f(B)$ <br> $f(B) = \begin{cases} \ln(B/B_c^*)/\ln(B_c/B_c^*), B \geq B_c^* \\ 0 \qquad\qquad\qquad\quad, else \end{cases}$ | $\sigma_s \sim -\frac{1}{\kappa^2}\left\{1 + \frac{\ln\left[1 - \left(\frac{\kappa B}{B_c}\right)^2\right]}{\left(\frac{\kappa B}{B_c}\right)^2}\right\}$ |
| σ vs temperature T | $\sigma_2(T) \sim \sigma_{20} + \Theta(T - T'^*) \cdot [\sigma_1(T) - \sigma_{20}]$ | $\sigma_s(T) \sim \Theta(T - T^*) \cdot (T - T^*)^\beta$ |

Table 1 summarizes, for bulk and surface, the effects as described before on the electrical conductivity σ of the nc electrons.

Analysing Fig. 1 (a) permits stating that the field dependent surface resistance factorizes into a temperature and field dependent part, as already published elsewhere [14]. This finding is important to discriminate models of different provenance. The field dependent part of the surface resistance, proportional to $Q^{-1}(B,T) - Q^{-1}_{max}(T)$, if plotted half-logarithmically vs. $T^{-1}$, follows very closely a common relation similar to the BCS surface resistance plus a residual part, in parallel lines depending on B. Its pre-factor A is linearly proportional to the residual surface resistance, for variable B, and logarithmically dependent on B. These observations confirm the ansatz as in eq. 16. As the surface resistance characterizing the individual loss mechanisms is additive, provided that the RF losses are smoothly distributed over the cavity surface, which we assume, the total surface resistance is composed of the contributions as in eqs. (3), (12) and (13), including the prescriptions of Table 1:

$$R_s = \left(A \cdot \frac{e^{-\Delta/T}}{T} + R_{res}\right) \cdot \{1 - f(B) + f(B) \cdot [\sigma_2/\sigma_1(T) + \Theta(T - T'^*) \cdot (1 - \sigma_2/\sigma_1(T))]\} + \{R_{s1} + R_{s2} \cdot$$

$$\Theta(T - T^*) \cdot (T - T^*)^\beta\} \cdot (-\kappa^{-2})\left\{1 + \frac{\ln\left[1 - \left(\frac{\kappa B}{B_c}\right)^2\right]}{\left(\frac{\kappa B}{B_c}\right)^2}\right\}. \quad (16)$$

The surface resistances $R_{s1}$ and $R_{s2}$ (not to be confused with the resistances $R_1$ and $R_2$, cf. eq. 6) are taken from ref. [9], eq. 48, and describe the non-temperature dependent and the temperature dependent contribution to the field-dependent RF losses at the surface. The first summand of eq. 16 describes RF losses in the bulk, the second one those at the surface.

The model as outlined so far is undoubtedly based on postulates, which are partly interrelated (such as the "percolation temperature", which depends on the percolation threshold for a specific arrangement of defects), and is therefore speculative to some extent. Hence checking with data will increase its credibility, as already done in ref. [9]. This is what follows now.

## *Discussion*

### *Data analysis*

Table 2: Result of least-square fitting for the data of Fig. 1(a) - left columns - and Fig. 1(b) - right columns.

| Data of Fig. 1 (a) | | Data of Fig. 1 (b) | |
|---|---|---|---|
| **Fit Parameter** | **Value** | **Value** | **Unit** |
| $\beta$ | $7 \pm^2_5$ | n/a | - |
| $R_{s1}$ | $3.5 \pm 3.5$ | $6.5 \pm 3.5$ | $n\Omega$ |
| $R_{s2}$ | $10 \pm 10$ | n/a | $n\Omega$ |
| $\sigma_2/\sigma_1(T)$ | $0.42 \pm 0.06$ | $0.38 \pm 0.12$ | - |
| $T^*$ | $1.3 \pm^{1.2}_{0.2}$ | n/a | K |
| $T'^*$ | $2.3 \pm 0.2$ | n/a | K |
| **Fixed Parameter** | **Value** | **Value** | **Unit** |
| A | $120 \cdot 10^3$ | $120 \cdot 10^3$ | $n\Omega \cdot K$ |
| $\Delta$ | 17.9 | 17.9 | K |
| $R_{res}$ | 2.0 | 1.8 | $n\Omega$ |
| $B_c^*$ | 9 | 9 | mT |
| $B_c$ | 190 | 190 | mT |
| $\kappa$ | 1 | 1 | - |
| n/a = not applicable because data do not allow observing a sudden change of slope Q(B) with T close $T^*$ and $T'^*$. | | | |

As to the fitting procedure, we made an effort to reduce the number of fit parameters to the utmost minimum by setting parameters as fixed, which were either known beforehand, such as $B_c$ and $\kappa$, which could be deduced from the low field Q-value, such as A, $\Delta$, and $R_{res}$, or which were derived from the low-field Q-increase, such as $B_c^*$. Consequently, only those parameters describing the low-field Q-increase, remained as adjustable parameters, such as $\beta$, $R_{s1}$, $R_{s2}$, $\sigma_2/\sigma_1(T)$, $T^*$, and $T'^*$, cf. Table 2.

We use the MATHEMATICA® package to find the best fit for the parameters, as summarized in Table 2, by minimizing the mean square error $\chi^2$,

$$\chi^2 = \{[R_{s,measured}(B,T) - R_s]/(RelativeError \cdot R_{s,measured}(B,T))\}^2 \qquad .$$

For the data of Fig. 1(a) a relative error for the measurement of the surface resistance of ±4 % results in $\chi^2 = 125$, close enough to the number of data points (167), reduced by the number of 6 fit parameters. This provides confidence into the model as described by eq. (16). For the data of Fig. 1(b) the relative error is larger (±10%) which leads to $\chi^2 = 129$ for a total number of data points of 127 and 2 fit parameters. Most of the fit parameters are uncorrelated and the $\chi^2$-function displays an inverse bell-shaped minimum. The correlated ones are either factors in a product or tied by a common B - or T - dependence, for which the correlation is understandable. We take as the error of the fit parameters the one where the $\chi^2$-value is twice its minimum value with all other fit parameters kept at their $\chi^2$-minimum value.

*Consistency check*

As a consistency check, we submitted data as shown in Table 2 to the proximity effect and percolation formalism as of ref. [9], cf. Table 3.

Table 3: Superconducting parameters of Nb and a compound of $Nb_x/N_{1-x}$.

|  | $(NV)_{S,N}$ | $\Theta_D$ [K] | $T_c$ [K] | $N_{S,N}$ [cm$^{-3}$] |
|---|---|---|---|---|
| Nb (S) | 0.283 [9] | 276 [9] | 9.25 [9] | 5.56·10$^{22}$ [9] |
| $Nb_x/N_{1-x}$ (N) | 0.196 | 174 [17] | 1.2 [18] | 6.46·10$^{22}$ [17] |
| "S" means strong superconductor (Nb), "N" means weak superconductor ($Nb_x/N_{1-x}$ compound) | | | | |

Some data for the $Nb_x/N_{1-x}$ compound are taken from ref. [17]. From the literature we know that the lowest critical temperature of the $Nb_x/N_{1-x}$ compound is 1.2 K [18], equivalent to 15 atomic percent of nitrogen in niobium (as obtained for a thin film in the cubic W-phase), hence x = 0.85. The electron-phonon coupling constant $(NV)_N$ for the $Nb_x/N_{1-x}$ compound at that lowest critical temperature of the $Nb_x/N_{1-x}$ compound is derived from the BCS-formula[b], with $\Theta_D$ being the Debye temperature,

$$T_c = 1.14 \cdot \Theta_D \cdot e^{-1/(NV)} \qquad , \qquad (17)$$

to $(NV)_N = 0.196$. The effective electron-phonon coupling constant $(NV)_{eff}$ for the $Nb_x/N_{1-x}$ compound embedded in the niobium is according to the proximity effect in the "Cooper-limit" [19, 20]

$$(NV)_{eff} = \frac{(NV)_N N_N v_N + (NV)_S N_S v_S}{N_N v_N + N_S v_S} \qquad , \qquad (18)$$

$v_N$ and $v_S$ being the volumes, $N_N$ and $N_S$ the electron densities, and $(NV)_N$ and $(NV)_S$ the superconducting coupling constants of the N and S components, respectively.

---

[b] As we interpolate the critical temperature of the composite between two experimentally known numbers (that of $Nb_{0.85}/N_{0.15}$ and that of Nb) there is no need to take into account the strong coupling theory.

Table 4: Separation of the contributions to the surface resistance of the weak sc defects related to the fit parameters of Figs. 1(a).

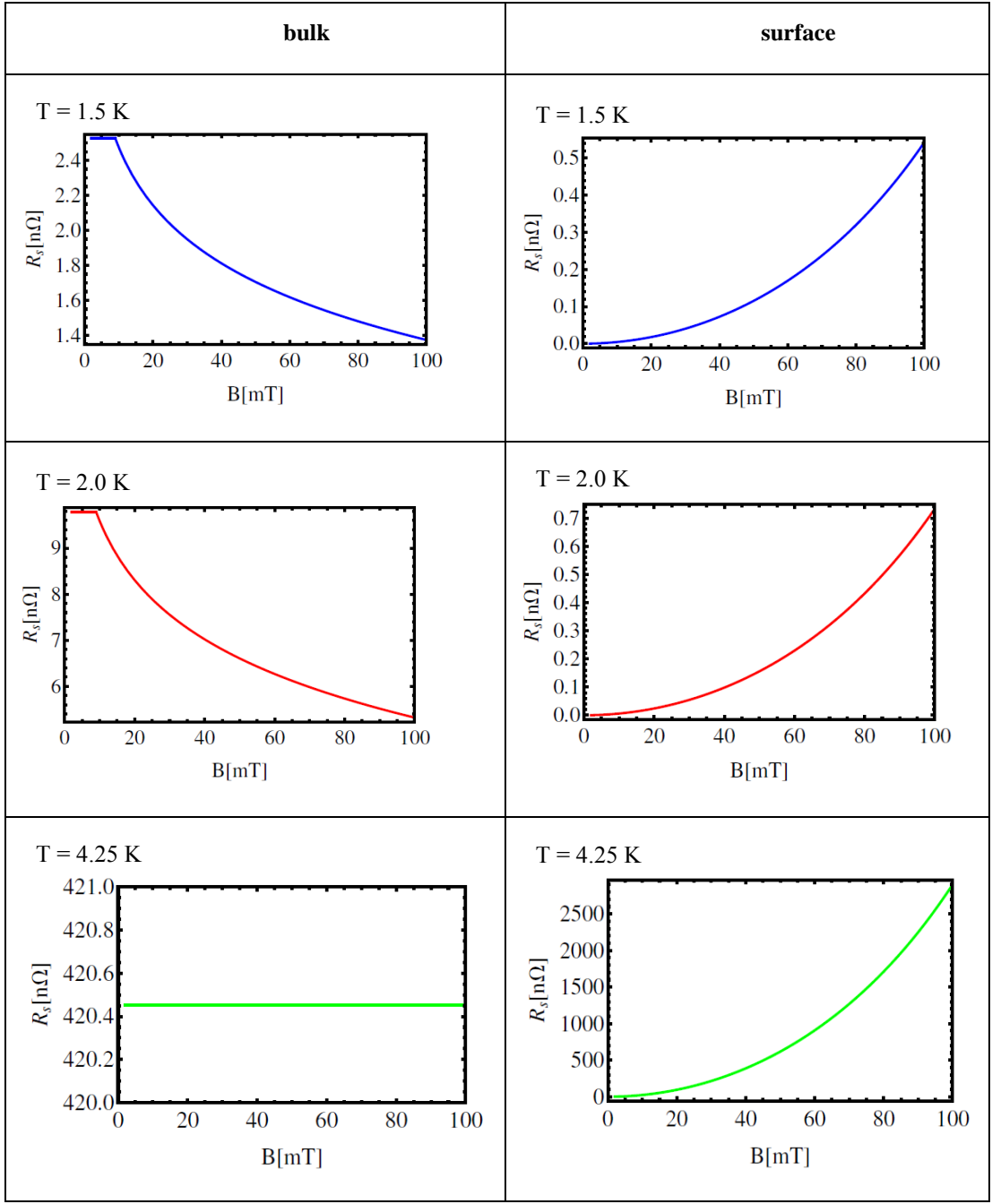

We know that the void percolation threshold for continuum percolation and a distribution of overlapping spheres (N) with equal radius and voids (S) in between is $x^* = v_S/(v_N+v_S) = 0.03$, above which there is long-range connectivity and therefore increased electrical conductivity. Applying this threshold for eq. 18 to determine the effective superconducting coupling constant $(NV)_{eff}$, and inserting it into eq. 17, with the Debye temperature averaged between that of niobium and that of the $Nb_x/N_{1-x}$ compound at $x^*$, we obtain a critical temperature for the composite of the $Nb_x/N_{1-x}$ compound embedded in the niobium matrix of $T_{c,eff} = 1.3$ K, close to the observed value for $T^*$. The threshold $x^* = 0.03$ corresponds to a volume fraction of $Nb_x/N_{1-x}$ to Nb of $v_N/v_S$

= 32. Practically this number describes overall coverage of the surface with $Nb_x/N_{1-x}$ enclosing 15 atomic percent nitrogen (x = 0.85).

We check now the consistency of this number with the SIMS depth profile of the nitrogen content (Fig. 2). With the atomic number of Nb (A = 93), its density ($\rho$ = 8.6 g·cm$^{-3}$), and Avogadro's number ($N_A$ = 6·10$^{23}$ Mol$^{-1}$), the atomic density of atoms at the surface is 5.5·10$^{22}$ cm$^{-3}$. For the relative composition N/Nb of the compound of 15 %, this number corresponds to 0.83·10$^{22}$ nitrogen atoms per cm$^3$, very close to what was found at the surface before removal of the uppermost layers by electro-polishing. Hence we conclude that, after electro-polishing, when the nitrogen volume density, 0.8·10$^{20}$ nitrogen atoms per cm$^3$, is lower by a factor of 100, as we learn from the SIMS depth profile, surface defects are still present, however dispersed by this same factor.

We also estimated the size of the compound inside the bulk from the Ginzburg-Landau parameter $\kappa \approx 1$ as of Table 2. The corresponding mean free path is 450 nm, with the London penetration depth $\lambda_L$ = 29 nm and the intrinsic coherence length $\xi_0$ = 33 nm. If the mean free path is determined by interstitially arranged nitrogen atoms, the nitrogen atomic volume density is 10$^{13}$ atoms·cm$^{-3}$. As the measured atomic nitrogen density is larger by a factor 10$^7$, the $Nb_x/N_{1-x}$ compound must have a linear dimension of about 215 atoms or 56 nm as upper limit (calculated from A, $\rho$ and $N_A$). This number is not in contradiction to the overall assumption which the described model is based on, namely that the defect size should be in the range of the coherence length of niobium. As we convinced ourselves, the coherence length $\xi_N$ of the compound induced by proximity from the Nb host metal lies in the range of the estimated defect size.

In Table 4 the contributions to the surface resistance $R_s$ of the weak sc defects are shown separately, as derived from the fit parameters (Table 2). The bath temperature increases downward the columns. On the left, the contribution to $R_s$ of the weak sc defects located in the bulk is shown, on the right that of the weak sc defects at the surface. From these plots the worst case is clearly identified as operation at 4.25 K and in presence of a large number of surface weak sc defects. On the contrary, the optimum case is at a temperature smaller than $T^{'*}$ (= 2.3 K) under toleration of weak sc defects in the bulk and the utmost lack of these at the surface.

*Conclusion*

We presented the essential features, partly as a summary of previously published work, of a model describing the field dependence of the Q-value (or, equivalently, the surface resistance) in sc bulk niobium accelerating cavities for the entire data range from 1.5 K to 4.25 K. We compared the model with recent data on so called "N-doping" of sc cavities. This model is based on the two-fluid description of the surface resistance and the postulated presence of weak sc defects. The model essentially uses a single parameter, the conductivities of the nc current components of the superconductor and of the weak sc defects when in the nc state. Other major features are the sc proximity effect, percolation behaviour, and the distinction between surface and bulk properties, the surface conditions being much more determinant to the power dissipation than the bulk conditions. The necessary conditions, according to this model, for obtaining a very small surface resistance (in the nΩ range) and an increase of the Q-value with field by "N-doping" are these: (i) minimization of the number of superficial weak sc defects by, e.g. electro-polishing, and their diffusion into the bulk by, e.g. thermal annealing; (ii) growing of weak sc defects in the bulk with large N content with the aim to push the percolation (fragmentation) to the upmost temperature, above the envisaged operation temperature; (iii) minimization of the number of weak sc defects in the bulk above the percolation temperature, possibly at the cost of weak sc defects with a larger N content below this temperature. The physical effects are, as to (i): avoiding the increase of nc carriers at the surface induced by the first entry of magnetic flux; and as to (ii) and (iii): impediment of the flow of nc current within the weak sc defects and hence reduction of the RF losses, if the defects have turned nc below the percolation temperature. The treatment should therefore aim at a homogeneous depth profile of $Nb_x/N_{1-x}$ compounds in the bulk, low in number, and non-existent or sufficiently deep inside the bulk if above the percolation temperature.

## ACKNOWLEDGEMENT

One of the authors (WW) likes to thank Professors Georg Hoffstaetter and Ralf Eichhorn for the invitation to visit Cornell's CLASSE Laboratory and for their hospitality during his stay.